\documentclass[%
reprint,
superscriptaddress,
showpacs,
 amsmath,amssymb,
 aps,
 prl,
]{revtex4-1}

\usepackage{graphicx}
\usepackage{dcolumn}
\usepackage{bm}
\usepackage{amsmath}
\usepackage{autobreak}
\usepackage{hyperref}
\usepackage[mathlines]{lineno}
\usepackage{natbib}
\usepackage{multirow}
\usepackage{booktabs}
\usepackage{xspace}
\usepackage{float}
\usepackage{diagbox}
\usepackage{subfigure}

\begin{document}

\preprint{APS/123-QED}

\title{Search for fractionally charged particles as lightly ionizing particles \\using public data from an underground direct detection experiment}

\author{Z.~H.~Zhang}
\email{Corresponding author: zhenhua@bnu.edu.cn}
\affiliation{Beijing Normal University at Zhuhai, Zhuhai 519087, China}
\author{J.~W.~Hu}
\email{Corresponding author: hujinwei@xndt.com.cn}
\affiliation{Beijing Honest Technology Co., Ltd., Beijing, 100083, China}
\author{X.~P.~Geng}
\affiliation{Institute of Applied Physics and Computational Mathematics, Beijing 100094, China}
\author{F.~L.~Liu}
\affiliation{China Institute of Atomic Energy, Beijing, 102413, China}
\affiliation{Center for Nuclear Study, The University of Tokyo, Wako-shi, 351-0198, Japan}
\author{Z.~H.~Zhang}
\affiliation{China Nuclear Power Technology Research Institute Co., Ltd., Shenzhen, 518000, China}

\date{\today}

\begin{abstract}
Approximately 60 years after the Milligan oil drop experiment in 1909, fractionally charged particles (FCPs) were reported to have been discovered in air-shower cores, and the search for FCPs continues to the present day in space, terrestrial and underground experiments. In this letter, relativistic FCPs are examined  as lightly ionizing particles in underground direct detection on the basis of electron recoil data from XENON-nT. For FCPs from supernova shocks, the best results are obtained via direct detection, excluding the FCP charge of $\left(3.6\times10^{-9} - 1.0\times10^{-2}\right)\cdot e$ corresponding to an FCP mass of 1 keV--556 GeV. For FCPs from general sources, the best limit is achieved, which is far ahead of other results in almost the entire parameter space, by excluding  the FCP flux of $\left(4.0\times10^{-4} - 4.0\times10^{-16}\right)$ cm$^{-2}\cdot$ s$^{-1}\cdot$ sr$^{-1}$ corresponding to the FCP charge of $\left(1\times10^{-6} - 1\right)\cdot e$.
\end{abstract}

\maketitle

\emph{Introduction.}
In 1909,  Millikan obtained a precision measurement of the charge of an isolated ion and denoted the test value of an elementary charge as  $e$~\cite{Millikan_1910, Millikan_1911}. However, electric charge quantization is currently a mystery that is unexplained by the standard model (SM) of particle physics. In 1969, McCusker and Cairns reported a track in a delayed-expansion Wilson cloud chamber produced by a fractionally charged particle (FCP, denoted ``$\mathrm{\chi}$''; also known as a ``milli- or minicharged particle") in an air-shower core~\cite{McCusker_1969}. Moreover, some difficult problems, such as dark matter (DM)~\cite{DMProblem_1987} and neutrino oscillations~\cite{NeutrinoOscillation_1998, NeutrinoOscillation_2002}, suggest that there must be some new physics beyond the SM.

Some researchers have made great efforts to study unconfined and free FCPs (in contrast to the quark's bound state) to extend the SM. Finding FCPs is helpful in testing some popular hypotheses, such as DM~\cite{Chun_1995, Kouvaris_2013, Shiu_2013, Muñoz_2018} and string theory~\cite{Schellekens_1990, Abel_2008, Shiu_2013}, as well expanding the SM. In this work, we search for FCPs as lightly ionizing particles (LIPs) with 1.16 ton$\cdot$year exposure data obtained from the XENON-nT experiment~\cite{XENONnT_ER2022} and update the limit on their parameter space.

Note that it is currently difficult to determine whether these FCPs are generated in the atmosphere or in outer space. Thus, we search for incident FCPs from general sources (including the atmosphere, the Sun and so on) in the view of the detector and mark them with the subscript ``gen''. Some studies have also attempted to generate FCPs via SM particle interactions in nuclear reactors~\cite{reactor_2007, TEXONO} and accelerators~\cite{Dobroliubov_1990, Perl_2009}.

In addition, FCPs in dark cosmic rays can be produced through supernova shock (SNS) and accelerated by the Fermi mechanism. Then, they can undergo radiative cooling, Coulomb scattering and energy redistribution during transport~\cite{Malkov_1998, Malkov_2016, Drury_2000, Berezhko_2006}. Finally, they reach the Earth, and their flux in the unit of $\mathrm{(GeV^{-1} \cdot cm^{-2} \cdot s^{-1} \cdot sr^{-1})}$ can be estimated via Eq.~\ref{eq::Flux}, which is derived from a comparison with protons~\cite{Hu_2017}.

\begin{equation}
    \left. \frac{\mathrm{d}\phi_{\mathrm{\chi}}}{\mathrm{d}E_{\mathrm{\chi}}} \right |_{\mathrm{SNS}}
    \approx 30 \cdot f_{\mathrm{\chi}}^{1.7} \cdot m_{\mathrm{\chi}}^{-1} \cdot E_{\mathrm{\chi}}^{-2.7}
    \label{eq::Flux}
\end{equation}
where $m_{\mathrm{\chi}}$ is the FCP mass in units of GeV; $E_{\mathrm{\chi}}$ is the FCP energy in units of GeV; $f_{\mathrm{\chi}}$ is defined as $f_{\mathrm{\chi}} \equiv Q_{\mathrm{\chi}}/e$; $Q_{\mathrm{\chi}}$ is the charge carried by an FCP; and $0<f_{\mathrm{\chi}}<1$. When $E_\mathrm{\chi} < \left(10^{18} \cdot f_\mathrm{\chi} \right)$ eV, the anisotropy of the FCP flux is not obvious. Impliedly, local DM density is 4.1 GeV/cm$^{-3}$, and the enrichment factor of FCP is 4.

\emph{FCP signal}
When FCPs reach the detector in direct detection experiments, there is a probability of interaction with the target atoms, resulting in a recoil energy of $E_\mathrm{r}$ for the atom electrons. The differential event rate of the interaction between FCPs and atoms in the detector is given by Eq.~\ref{eq::Rr-Er}.

\begin{equation}
    \left . \frac{\mathrm{d}R_\mathrm{r}}{\mathrm{d}E_\mathrm{r}} \right |_{\mathrm{SNS}} = \rho_A \cdot \int^{E_\mathrm{max}}_{E_\mathrm{min}}
    \frac{\mathrm{d}\sigma_{\mathrm{\chi} A}}{\mathrm{d}E_\mathrm{r}} \cdot \frac{\mathrm{d}\phi_\mathrm{\chi}}{\mathrm{d}E_\mathrm{\chi}} \cdot \mathrm{d}E_\mathrm{\chi}
    \label{eq::Rr-Er}
\end{equation}
where $\rho_A$ is the number density of the target atoms and $\frac{\mathrm{d}\sigma_{\mathrm{\chi} A}}{\mathrm{d}E_\mathrm{r}}$ is the differential cross-section between the FCP and the atom. The number density of xenon atoms is $\rho_A = 4.58\times10^{27}$ ton$^{-1}$, and $\frac{\mathrm{d}\sigma_{\mathrm{\chi} A}}{\mathrm{d}E_\mathrm{r}}$ is estimated with the equivalent photon approximation~\cite{Weizsäcker_1934, Williams_1934}, as shown in Eq.~\ref{eq::sigma-E}. Focusing on the relativistic FCPs from the SNS, we set $E_\mathrm{min} = \left( 10\cdot m_\mathrm{\chi} \right)$ and $E_\mathrm{max} = \left(10^6 \cdot f_\mathrm{\chi}\right)$ GeV.

\begin{equation}
    \left . \frac{\mathrm{d}\sigma_{\mathrm{\chi} A}}{\mathrm{d}E_\mathrm{r}} \right |_{\mathrm{SNS}}
    = f_\mathrm{\chi}^2 \cdot \frac{2\alpha}{\pi} \cdot \frac{\sigma_\mathrm{\gamma}(E_\mathrm{r})}{E_\mathrm{r}} \cdot \mathrm{ln} \left(\frac{E_\mathrm{\chi}}{m_\mathrm{\chi}} \right)
    \label{eq::sigma-E}
\end{equation}
where the fine structure constant is $\alpha = 1/137$ and $\sigma_\mathrm{\gamma}(E_\mathrm{r})$ is the photoelectric absorption cross-section of the xenon atom obtained from the Evaluated Nuclear Data File~\cite{ENDF_Website}. The FCP is considered a LIP, which causes less ionization in the detector because $\frac{\mathrm{d}\sigma_{\mathrm{\chi} A}}{\mathrm{d}E_\mathrm{r}}$ is proportional to $f_\mathrm{\chi}^2$.

\begin{figure}[!htbp]
\centering 	
\includegraphics[width=0.48\textwidth]{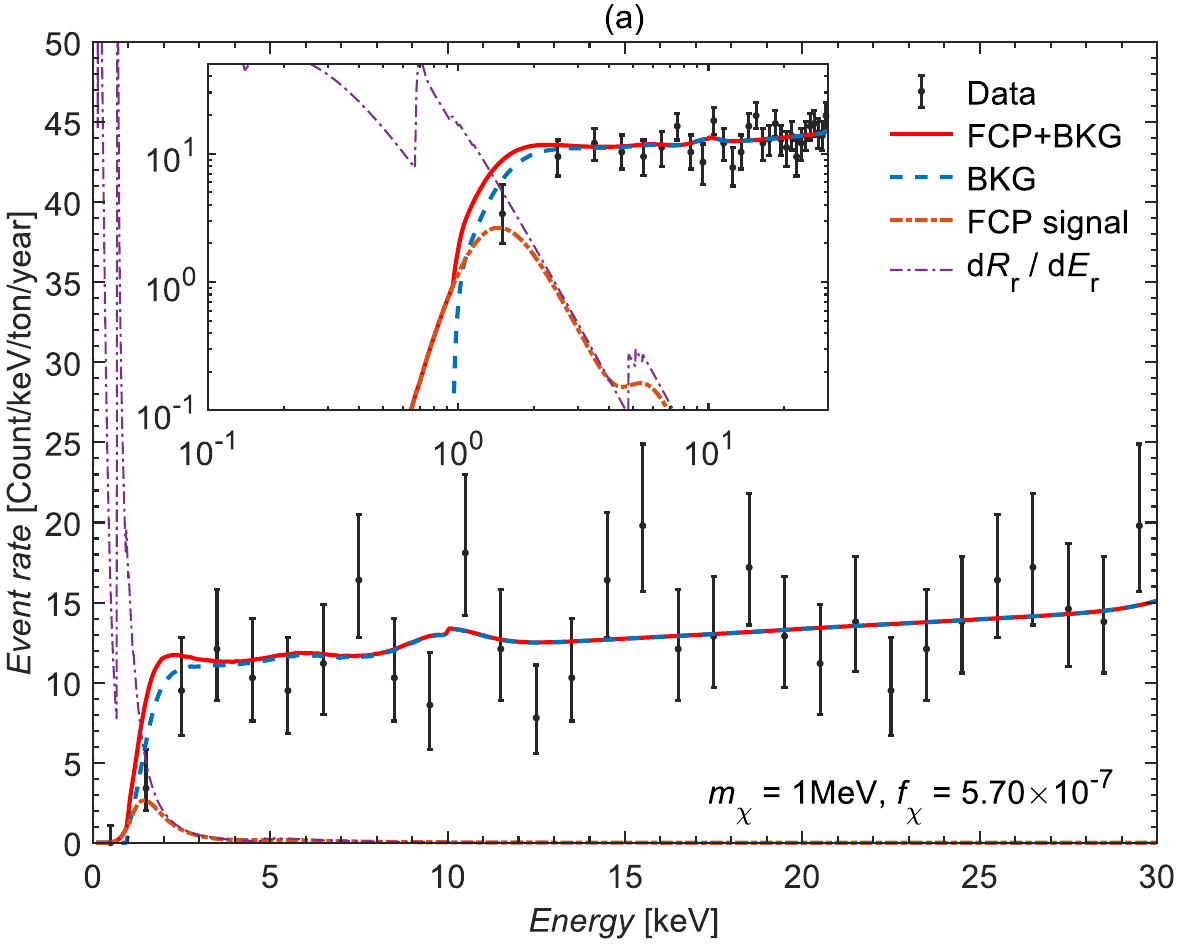}      \includegraphics[width=0.48\textwidth]{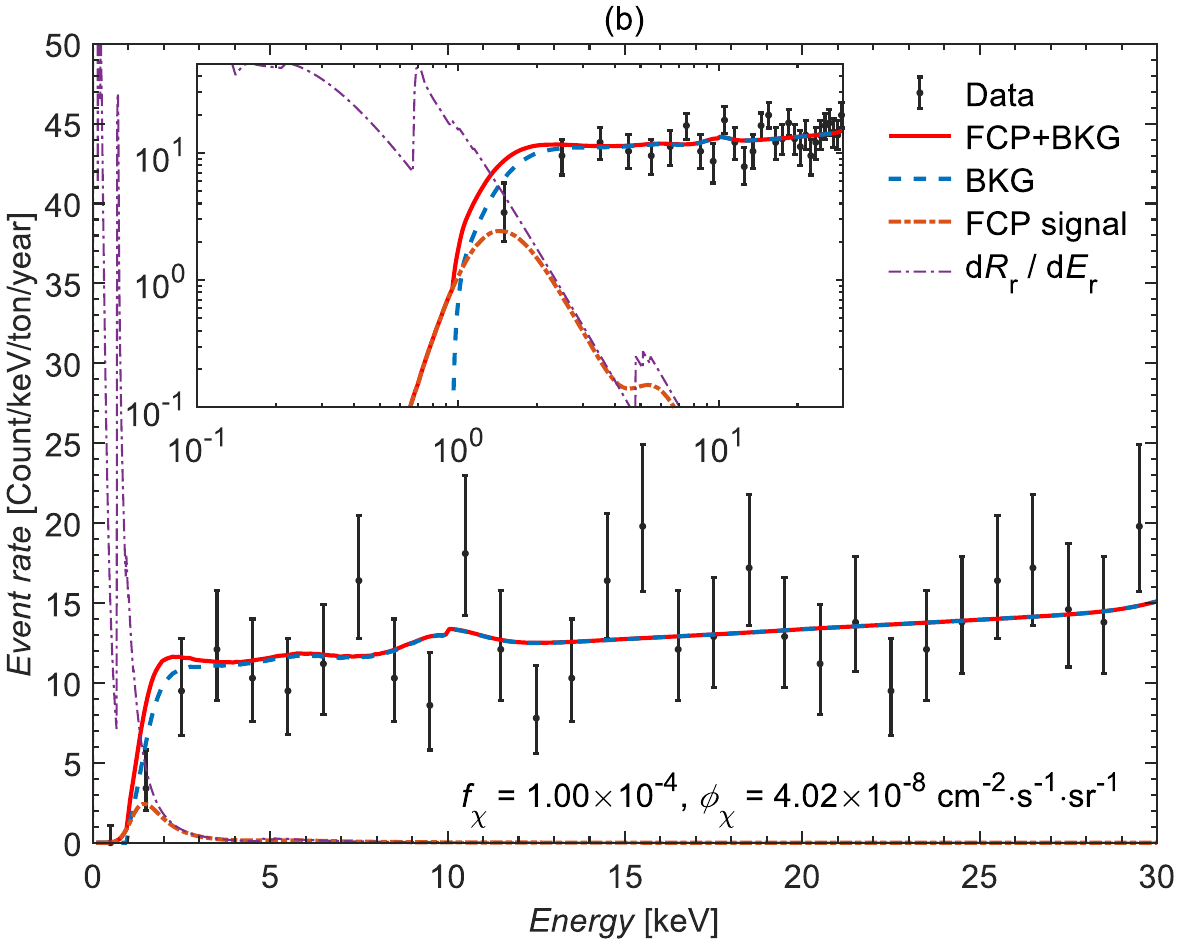}  	\caption{Matching between the experimental and expected spectra in the case of FCPs from (a) SNS and (b) general sources. The data points with error bars represent the electron recoil spectrum in the XENON-nT experiment, which are obtained from Fig. 5 in Ref.~\cite{XENONnT_ER2022}. The orange dotted-dashed line (FCP signal) and the blue dashed line (BKG) are combined to yield the red solid line (FCP+BKG). BKG is the background $B_0$ in Ref.~\cite{XENONnT_ER2022} and the website~\cite{XENON_Website}. The value $\frac{\mathrm{d}R_\mathrm{r}}{\mathrm{d}E_\mathrm{r}}$ before considering the energy resolution and efficiency is plotted with a thin purple dotted-dashed line.}
\label{fig::SpecMatching}
\end{figure}

As shown in Eq.~\ref{eq::Re-Ee}, the expected signal of FCPs $\frac{\mathrm{d}R_\mathrm{e}}{\mathrm{d}E_\mathrm{e}}$ is derived by convolving $\frac{\mathrm{d}R_\mathrm{r}}{\mathrm{d}E_\mathrm{r}}$ over the energy resolution of the detector and multiplying it by the total efficiency of the experiment. The difference between these signals is clearly indicated in Fig.~\ref{fig::SpecMatching}. After considering the energy resolution and efficiency, the curve becomes smooth and decreases, especially in the lower-energy region. This implies that a detector with excellent energy resolution and a low energy threshold performs better at searching for FCPs.

\begin{equation}
    \frac{\mathrm{d}R_\mathrm{e}}{\mathrm{d}E_\mathrm{e}} = \frac{\mathrm{d}R_\mathrm{r}}{\mathrm{d}E_\mathrm{r}}
        * \sigma_\mathrm{res} \cdot \eta_\mathrm{eff}
    \label{eq::Re-Ee}
\end{equation}
where the energy resolution $\sigma_\mathrm{res} \mathrm{(keV)} = 0.310 \cdot \sqrt{E \mathrm{(keV)}} +0.0037$ is referenced from XENON-1T~\cite{XENON1T_ER2020, XENONnT_ER2022} and the total efficiency $\eta_\mathrm{eff}$ is obtained from public data on XENON's website~\cite{XENON_Website}.

From the perspective of the detector, the incident FCPs from general sources have an undefined distribution of mass and energy because of their diverse sources. In this letter, we use a conservative estimation, and Eq.~\ref{eq::Rr-Er} is simplified to Eq.~\ref{eq::Rr-flux}. Moreover, $\frac{E_{\mathrm{\chi}}}{m_{\mathrm{\chi}}}$ in Eq.~\ref{eq::sigma-E} is set to 10 here.

\begin{equation}
    \left . \frac{\mathrm{d}R_\mathrm{r}}{\mathrm{d}E_\mathrm{r}} \right |_{\mathrm{gen}} = \rho_A \cdot \phi_\mathrm{\chi}
    \cdot \frac{\mathrm{d}\sigma_{\mathrm{\chi} A}}{\mathrm{d}E_\mathrm{r}}
    \label{eq::Rr-flux}
\end{equation}

\begin{figure}[!htbp]
\centering
\includegraphics[width=0.48\textwidth]{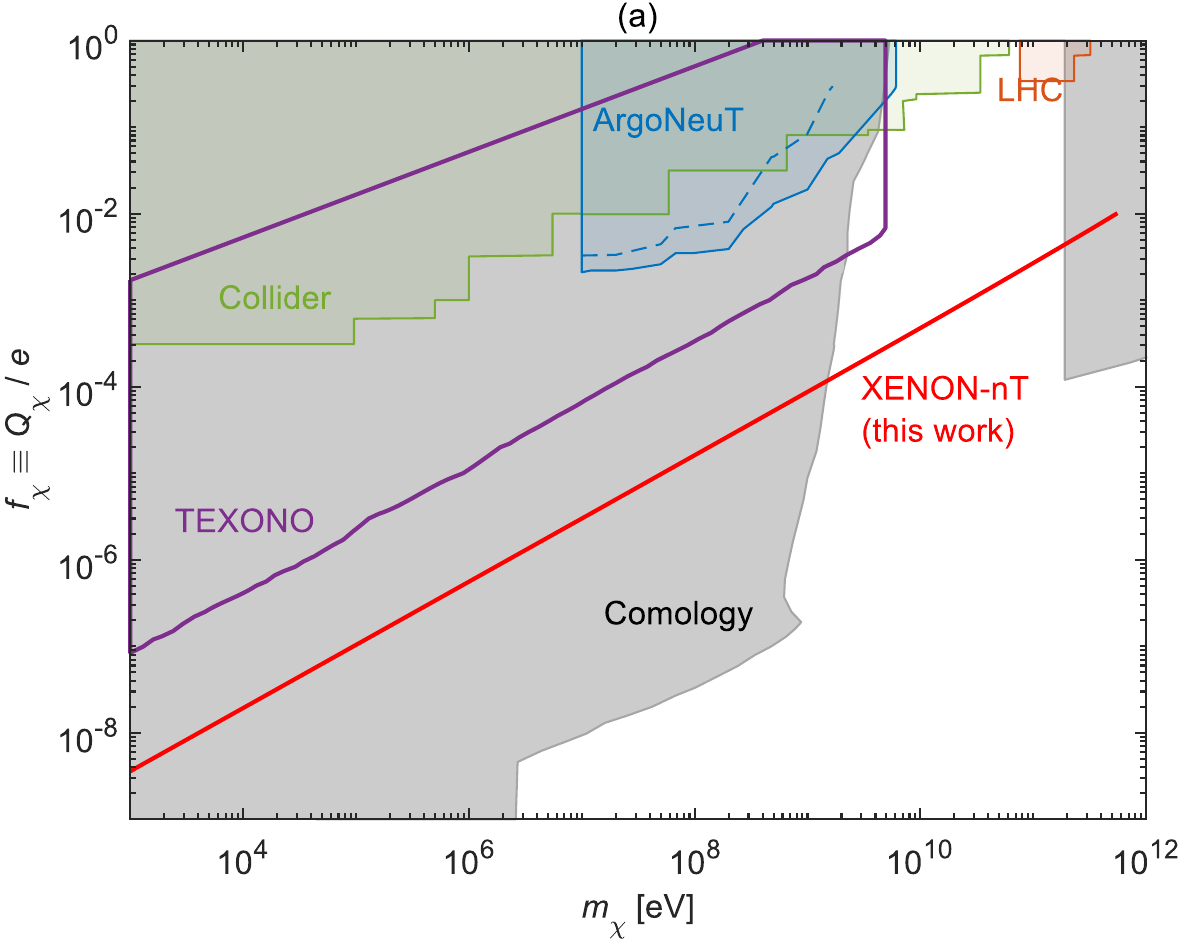} 
\includegraphics[width=0.48\textwidth]{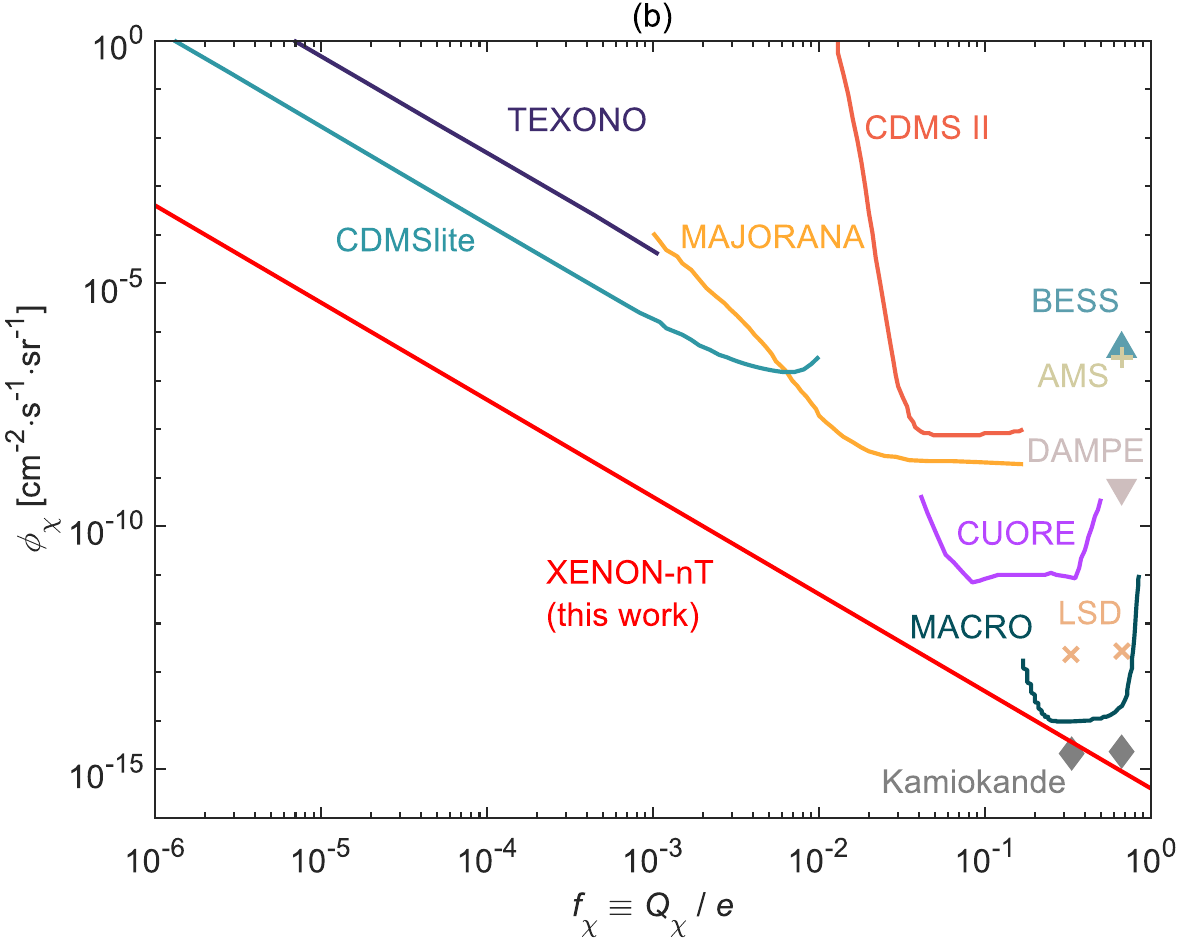} 
\caption{The limits on FCPs from (a) SNS  and (b) general sources at the 90\% one-sided confidence level. In the top panel (a), the constraint on $f_\mathrm{\chi}$ as a function of the $m_\mathrm{\chi}$ of FCPs from SNS is plotted as a red line. The summary limit of cosmology refers to the lower outline of the cosmology results (such as star~\cite{star1, star2}, supernova~\cite{SN}, BBN~\cite{Vogel_2014}, CMB~\cite{CMB, Vogel_2014} and DM~\cite{DM}) in the top panel in Fig. 1 in Ref.~\cite{Cosmology_2016} and is shown in gray shading. The results from collider~\cite{Collider1, Collider2}, LHC~\cite{LHC} and ArgoNeuT~\cite{ArgoNeuT} are shown in green, orange and blue shading, respectively. The limit obtained from TEXONO~\cite{TEXONO} is plotted as a purple line. In the bottom panel (b), the constraint on $\phi_\mathrm{\chi}$ as a function of $f_\mathrm{\chi}$ of FCPs from general sources is plotted as a red line. The limits obtained from space experiments (such as BESS~\cite{BESS}, AMS~\cite{AMS01} and DAMPE~\cite{DAMPE}), terrestrial experiments (such as TEXONO~\cite{TEXONO}) and other underground experiments (such as LSD~\cite{LSD}, Kamiokande~\cite{Kamiokande}, MACRO~\cite{MACRO}, CDMS~\cite{CDMSlite}, MAJORANA~\cite{MAJORANA} and CUORE~\cite{CUORE_2024}) are also shown for comparison.}
\label{fig::limits} 
\end{figure}

\emph{Results and discussion}
As shown in Fig.~\ref{fig::SpecMatching}, the FCP signal (dotted-dashed line) and background (dashed line) contribute to the expected energy spectrum (solid line). The matching between the expected spectrum and the measured spectrum is obtained via minimum-$\chi^2$ analysis, as shown in Eq.~\ref{eq::eq4}. The limits on FCPs from SNS and general sources are plotted in Fig.~\ref{fig::limits}(a) and (b), respectively.

\begin{equation}
\chi^2 = \sum_i \frac{\left[n_i-\left(S_i+B_i\right)\right]^2}{\Delta^2_i},
\label{eq::eq4}
\end{equation}
where $n_i$ and $\Delta _i$ are the experimental data and standard deviation with statistical and systematical components in the $i$-th energy bin extracted from Fig. 5 in Ref.~\cite{XENONnT_ER2022}, respectively; $S_i$ is the FCP event rate in the $i$-th energy bin; and $B_i$ is the background contribution in the $i$-th energy bin, obtained from Ref.~\cite{XENONnT_ER2022, XENON_Website}.

We find that the no-signal hypothesis yields $\chi^2_0 = 15.91$. On the basis of the Feldman--Cousins unified approach~\cite{FCChiSquare}, the limit of the 90\% one-sided confidence level is derived with $\chi^2 = \chi^2_0+2.71$.

The constraint on $f_\mathrm{\chi}$ as a function of the $m_\mathrm{\chi}$ of FCPs from SNS obtained in this work is plotted as a red line in the top panel in Fig.~\ref{fig::limits}. We use a cutoff of $f_\mathrm{\chi} = 0.01$ because an FCP with $f_\mathrm{\chi} > 0.01$ would behave like a muon, and the attenuation in the atmosphere and rock cannot be ignored. The results from cosmology, colliders, the LHC, ArgoNeuT and TEXONO are also plotted for comparison. This work updates the limits obtained from direct detection and accelerator experiments and fills the gap in cosmology observations for FCP masses ranging from 2 GeV to 200 GeV.

TEXONO and XENON-nT are both direct detection experiments for FCPs. In TEXONO, the BKG is $\sim10$ Count/keV/kg/day, and the energy threshold is 300 eV. In XENON-NT, the BKG is $\sim10$ Count/keV/ton/yr, and the energy threshold is 1 keV. The background of TEXONO is 365000 times that of XENON-nT, but the result obtained by TEXONO is only 25 times that of XENON-nT instead of 600 times. As discussed in the last section, the energy threshold and energy resolution are also two important factors in determining sensitivity to FCPs. Notably, the 2-hit result (blue shading) of ArgoNeuT is significantly better than the 1-hit result (blue dashed line). Both XENON-nT and ArgoNeuT are time-projection chambers; the travel length of the FCP is $\sim0.9$ m in the ArgoNeuT detector but $\sim1.6$ m in the XENON-nT detector ~\cite{XENONnT_EPJC2024}. If the FCP is further considered to deposit energy multiple times in XENON-nT, the limit result will be more stringent. 

The constraint on $\phi_\mathrm{\chi}$ as a function of the $f_\mathrm{\chi}$ of FCPs from general sources is plotted as a red line in the bottom panel of Fig.~\ref{fig::limits}. The results from space observations (BESS~\cite{BESS}, AMS~\cite{AMS01} and DAMPE~\cite{DAMPE}) and direct detection (LSD~\cite{LSD}, Kamlikande~\cite{Kamiokande}, MACRO~\cite{MACRO}, CDMS~\cite{CDMSII, CDMSlite}, MAJORANA~\cite{MAJORANA}, TEXONO~\cite{TEXONO} and CUORE~\cite{CUORE_2024}) are also plotted for comparison. The most stringent limit is achieved, covering almost all of the $f_\mathrm{\chi}$ range except $f_\mathrm{\chi} = 1/3$.

In this work, the limits on $\left(m_\mathrm{\chi}, f_\mathrm{\chi}\right)$ for FCPs from SNS and $\left(f_\mathrm{\chi}, \phi_\mathrm{\chi}\right)$ for FCPs from general sources are derived, and the existing results are updated. In the future, as the radiation background and energy threshold are further reduced, the energy resolution is further improved, and the exposure is further increased, the search for FCPs and new physics beyond the SM in underground direct detection will continue play an advanced role with space and terrestrial experiments together.

\bibliography{LIPs}

\end{document}